\documentclass[english,10pt, conference, letterpaper]{IEEEtran}
\usepackage[T1]{fontenc}
\usepackage[latin9]{inputenc}
\usepackage{units}
\usepackage{algorithm2e}
\usepackage{amsmath}
\usepackage{amsthm}
\usepackage{amssymb}
\usepackage{graphicx}

\makeatletter
\theoremstyle{plain}
\newtheorem{thm}{\protect\theoremname}
\theoremstyle{definition}
\newtheorem{defn}[thm]{\protect\definitionname}
\theoremstyle{plain}
\newtheorem{cor}[thm]{\protect\corollaryname}
\theoremstyle{plain}
\newtheorem{lem}[thm]{\protect\lemmaname}

\IEEEoverridecommandlockouts

\usepackage[font=small]{caption}
\usepackage{bbm}
\RestyleAlgo{ruled}

\newcommand{\Tau}{\mathrm{T}}

\author{
\IEEEauthorblockN{Jason Cloud}
\IEEEauthorblockA{Dolby Laboratories\\
San Francisco, CA USA\\
email: jason.cloud@dolby.com}
\and
\IEEEauthorblockN{Muriel M\'{e}dard}
\IEEEauthorblockA{Research Laboratory of Electronics\\
Massachusetts Institute of Technology\\
Cambridge, MA USA\\
email: medard@mit.edu}
}

\IEEEaftertitletext{\vspace{-0.75cm}}

\makeatother

\usepackage{babel}
\providecommand{\corollaryname}{Corollary}
\providecommand{\definitionname}{Definition}
\providecommand{\lemmaname}{Lemma}
\providecommand{\theoremname}{Theorem}

\begin{document}

\title{Multi-Path Low Delay Network Codes}
\maketitle
\begin{abstract}
The capability of mobile devices to use multiple interfaces to support
a single session is becoming more prevalent. Prime examples include
the desire to implement WiFi offloading and the introduction of 5G.
Furthermore, an increasing fraction of Internet traffic is becoming
delay sensitive. These two trends drive the need to investigate methods
that enable communication over multiple parallel heterogeneous networks,
while also ensuring that delay constraints are met. This paper approaches
these challenges using a multi-path streaming code that uses forward
error correction to reduce the in-order delivery delay of packets
in networks with poor link quality and transient connectivity. A simple
analysis is developed that provides a good approximation of the in-order
delivery delay. Furthermore, numerical results help show that the
delay penalty of communicating over multiple paths is insignificant
when considering the potential throughput gains obtained through the
fusion of multiple networks.
\end{abstract}

\section{Introduction}

The increasing availability of wireless devices with multiple radios
is driving the push to merge network resources across multiple radio
technologies and cellular access nodes. Prime examples include the
desire to offload traffic from cellular networks to WiFi networks
and the desire to simultaneously utilize both macro and small cells
in 5G networks. While the merging of network resources has the potential
to drastically increase throughput, packet losses due to congestion,
poor link quality, transient network connections, etc. can have serious
consequences for meeting users' quality of service (QoS) requirements.
A multi-path streaming code, derived from a low delay code designed
for single paths \cite{Karzand_2015}, is presented that helps overcome
these challenges by reducing the end-to-end, or in-order delivery,
delay. Through an analysis of the in-order delivery delay, we further
show that merging parallel networks together using network coding
enables the summation of individual path throughputs without significant
impacts to the overall in-order delivery delay. 

The delivery of information in the order it was first transmitted
is a requirement for most applications. Unfortunately, packet losses
that occur during transmission can cause significant disruptions and
delays. As an example, automatic repeat request (ARQ) is one approach
to recover from these packet losses. Whenever a packet loss occurs,
all packets received after the loss are buffered until ARQ corrects
the erasure. This takes on the order of a round-trip time ($RTT$)
or more. If the $RTT$ (more precisely the bandwidth-delay product
($BDP$)) is small, the disruption in packet delivery is relatively
minor. However if the $BDP$ is large, the added delay necessary to
recover from a packet erasure can be detrimental to the QoS of delay
sensitive applications.

Forward error correction (FEC) is one method to help minimize the
disruptions created by packet erasures. Reducing delay through the
use of FEC has been a topic of interest that has gained popularity
in the past few years. The delay performance of generation, or block,
based codes were investigated by \cite{Cloud_15}. In addition, the
delay gains of streaming codes were investigated by \cite{Karzand_2015,joshi_playback_2012,toemoeskoezi_delay_2014}.
While each of these references show that coding outperforms non-coding
approaches (e.g., ARQ) in terms of reducing delay, they focused on
the case where a single path is used for transmission. The extension
of these schemes to the case where communication occurs over multiple
parallel networks with different packet loss rates, transmission rates,
and propagation delays has the potential to realize more gains. Not
only will coding help recover from packet losses quickly, but it can
also reduce the complexity of scheduling packet retransmissions across
the different network paths.

The study of delay performance for coded schemes in multi-path environments
has been somewhat limited. The delay-rate trade-off with various multi-path
routing and coding approaches were investigated in \cite{Han_2008,Ronasi_2011}.
Multiple description coding and layered coding over multiple paths
was looked at by \cite{Nguyen_2004}. Of primary note is the work
by \cite{GarciaSaavedra_2015}. They propose an algorithm called Stochastic
Earliest Delivery Path First (S-EDPF) that combines packet scheduling
with a coding approach that is similar to the one presented within
this paper. While this is the case, there are some notable differences.
First, their work assumes redundancy is only transmitted on a single
path while the algorithm presented here is general enough to allow
redundancy to be transmitted on multiple paths. Second, we provide
a closed-form, straightforward analysis of the in-order delivery delay,
while the complex analytical model in \cite{GarciaSaavedra_2015}
that uses binomial distributions requires several assumptions and
relaxations. Finally, the rate-delay trade-off and a comparison illustrating
the delay penalty between single-path and multi-path transport is
shown while \cite{GarciaSaavedra_2015} does not.

The remainder of this paper is organized as follows. Section \ref{sec:Multi-Path-Streaming-Code}
describes the multi-path streaming code. This includes a discussion
regarding the code rate used on each path and the management of the
code window used to generate redundancy. Section \ref{sec:System-Model}
describes the system model that is used for the analysis of the in-order
delivery delay presented in Section \ref{sec:Analysis}. A comparison
of the analysis with simulated results in shown in Section \ref{sec:Numerical-and-Simulation}.
This section demonstrates that the analysis provides a good estimate
of the in-order delivery delay, the use of network coding to merge
parallel network paths results in gains to throughput without impacting
the delay, and the trade-off between rate and delay. Finally, the
paper is concluded in Section \ref{sec:Conclusions}.

\section{Multi-Path Streaming Code Algorithm\label{sec:Multi-Path-Streaming-Code}}

Consider a systematic coding scheme based on random linear network
coding (RNLC) \cite{ho_random_2006} that allows a server to communicate
over parallel paths or networks while helping to reduce the overall
in-order delivery delay. Information packets $\boldsymbol{p}_{i},i\in\left[1,M\right]$,
are injected into each network uncoded. Note that the server has limited
knowledge of the packets that will be sent in the future (i.e., it
does not have access to the entire file). If an opportunity arises
that allows the server to transmit a new packet, it does so without
attempting to ensure specific packets arrive at the client in a predetermined
order. After a specific number of information packets have been transmitted
on any given path, the server generates and transmits a coded packet
$\boldsymbol{c}_{i}$ on a path of its choosing to help overcome any
packet losses that may have occurred.
\begin{figure}
\begin{centering}
\includegraphics[width=0.9\columnwidth]{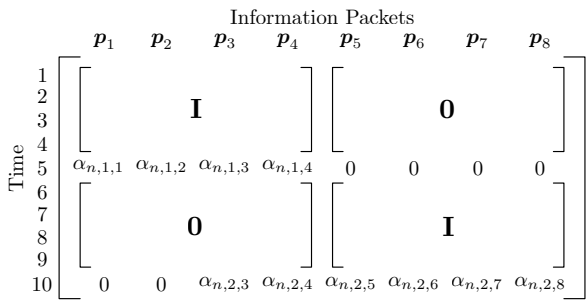}
\par\end{centering}
\caption{Example generator matrix used to produce the streaming code. The elements
of the matrix contain the coefficients used to produce each transmitted
packet. The rows show the composition of each transmitted packet and
the columns indicate the information packet that must be transferred.\label{fig:Example-Sliding-Window}}
\end{figure}

Define $l_{i}$ to be the duration between transmitted coded packets
on path $i\in\mathcal{P}$ (i.e., $l_{i}-1$ information packets are
transmitted followed by a single coded packet). This results in a
code rate of $c_{i}=\nicefrac{l_{i}-1}{l_{i}}$. If a path is idle,
the server will transmit either an information packet or coded packet
depending on the previously transmitted packets on that specific path.
When a coded packet is generated, the information packets used to
produce the linear combination are drawn from a dynamically changing
code window defined by the 2-tuple $w=\left(w_{L},w_{U}\right)$.
This results in the following packet:

\begin{equation}
\boldsymbol{c}_{n,k}=\sum_{i=w_{L}}^{w_{U}}\alpha_{n,k,i}\boldsymbol{p}_{i}.\label{eq:CodedPacket}
\end{equation}
The coefficients $\alpha_{n,k,i}\in\mathbb{F}_{q}$ are chosen at
random and each information packet $\boldsymbol{p}_{i}$ is treated
as a vector in $\mathbb{F}_{q}$. All of this is summarized in Algorithm
\ref{alg:Multi-path-coding-algorithm} where $\mathbbm{1}_{i}$ is
a vector of size $i$ consisting of all ones. In addition, an example
of the generator matrix used to produce the streaming code is provided
in Figure \ref{fig:Example-Sliding-Window}. In the example, information
packets $\boldsymbol{p}_{1}$ through $\boldsymbol{p}_{4}$ and $\boldsymbol{p}_{5}$
through $\boldsymbol{p}_{8}$ are transmitted systematically in time-slots
$1$ through $4$ and $6$ through $9$ respectively. In time-slots
$5$ and $10$, coded packets $\boldsymbol{c}_{n,1}=\sum_{i=1}^{4}\alpha_{n,1,i}\boldsymbol{p}_{i}$
and $\boldsymbol{c}_{n,2}=\sum_{i=3}^{8}\alpha_{n,2,i}\boldsymbol{p}_{i}$
are transmitted respectively. It is assumed in this example that the
server has obtained feedback from the client by time $10$ indicating
that it successfully received and decoded packets $\boldsymbol{p}_{1}$
and $\boldsymbol{p}_{2}$. This allows the server to adjust the lower
edge of the coding window to exclude the packets during the generation
of coded packet $\boldsymbol{c}_{n,2}$.
\begin{algorithm}
\SetAlgoNoLine 
\SetAlgoNoEnd 
\DontPrintSemicolon 
\SetKwFor{ForEach}{for each}{do}{endfch} 
Initialize $k=1$ and $\underbar{u}=\mathbbm{1}_{|\mathcal{P}|}$\; 
\While{$k\leq M$}{ $n\leftarrow$ First idle path found\; 
\If{$u_n < l_n$}{ Transmit packet $\boldsymbol{p}_k$\; $u_n\leftarrow u_n +1$\; $k\leftarrow k+1$\; } 
\Else{ Transmit coded packet $\boldsymbol{c}_{n,k}=\sum_{i=w_L}^{w_U} \alpha_{n,k,i} \boldsymbol{p}_i$\; $u_n\leftarrow 1$\; } }

\caption{Streaming Multi-Path Code Generation\label{alg:Multi-path-coding-algorithm}}
\end{algorithm}

Before proceeding, it must be noted that Algorithm \ref{alg:Multi-path-coding-algorithm}
does not explicitly take advantage of feedback in determining when
to inject coded packets into the packet stream. Rather, feedback is
only used to estimate the packet erasure probability $\epsilon_{i}$
on each path $i\in\mathcal{P}$. This is done in order to simplify
the analysis that will be presented later. However, using feedback
can only improve the algorithm's performance; and implemented versions
should use feedback intelligently when determining when to inject
redundancy to help reduce the delay further.

While Algorithm \ref{alg:Multi-path-coding-algorithm} is fairly simple,
two topics jump out that require special consideration. First, the
selection of code rates $c_{i}$ on every path $i\in\mathcal{P}$
must be done properly to ensure that the client can decode within
a reasonable time period regardless of the observed packet losses.
Second, management of the coding window $w$ must be performed carefully
to ensure coded packets add to the knowledge space of the client.
These two topics will be addressed by defining the following:
\begin{defn}
\label{def:Coding-Policy}A coding policy $\pi$ determines the code
structure and rates used on each path between a server and client.
\end{defn}
In other words, the coding policy defines the code rates used to generate
coded packets on each path, as well as the algorithm for managing
$w$. There are, in fact, an infinite number of coding policies. However,
policies that allow the client to decode with high probability within
a reasonable time frame are the only ones of interest. This leads
to the next definition:
\begin{defn}
\label{def:A-coding-policyAdmissibleCodingPolicy}A coding policy
that ensures the client will decode with probability equal to $1$
is said to be admissible.
\end{defn}
Let $\prod=\left\{ \pi_{1},\pi_{2},\pi_{3},\ldots\right\} $ be the
set of all admissible coding policies. The code rates and code window
management rules for policy $\pi_{i}$ will be referred to as the
$\left|\mathcal{P}\right|$-tuple $c\left(\pi_{i}\right)=\left(c_{1}\left(\pi_{i}\right),\ldots,c_{\left|\mathcal{P}\right|}\left(\pi_{i}\right)\right)$
and ${\cal W}\left(\pi_{i}\right)$ respectively. The following sub-sections
will help define both $c\left(\pi_{i}\right)$ and ${\cal W}\left(\pi_{i}\right)$
for each policy $\pi\in\Pi$.

\subsection{Code Rate Selection\label{subsec:Code-Rate-Selection}}

An admissible coding policy must ensure the client's capability to
decode. One of the most important parts is choosing the appropriate
code rate $c\left(\pi_{i}\right)$. The following theorem helps determine
this rate where its proof is provide in the appendix.
\begin{thm}
\label{thm:AdmissibleCodingPolicyConstraints}An admissible coding
policy $\pi$ must satisfy the following constraints:
\begin{equation}
\sum_{i\in\mathcal{P}}\left(1-\epsilon_{i}\right)\left(\left(1-\epsilon_{i}\right)-c_{i}\left(\pi\right)\right)r_{i}>0,\label{eq:admissable_policies-1}
\end{equation}
and
\begin{equation}
c_{i}\left(\pi\right)\in\left[0,1\right].\label{eq:admissable_policies-2}
\end{equation}
 
\end{thm}
Transmitting coded packets over multiple networks maybe the appropriate
strategy in some cases; but in others, it maybe better to only send
coded packets over a single network. This leads to the following corollary.
\begin{cor}
\label{cor:For-the-caseAdmissibleCodingPolicyConstraints-SinglePath}For
the case when coded packets are only transmitted over a single path
and there exists a $\pi$ such that $c_{i}\left(\pi\right)\in\left[0,1\right]$
and $c_{j}\left(\pi\right)=1$ for $i,j\in\left[1,\left|\mathcal{P}\right|\right]$,
$i\neq j$ , the admissible coding policy $\pi$ must satisfy the
following:
\end{cor}
\begin{eqnarray}
c_{i}\left(\pi\right) & < & 1-\frac{\sum_{k\in\mathcal{P}}\left(1-\epsilon_{k}\right)\epsilon_{k}r_{k}}{\left(1-\epsilon_{i}\right)r_{i}}.\label{eq:admissable_policy_single_path}
\end{eqnarray}

\subsection{Code Window Management\label{subsec:Coding-Window-Management}}

For admissible coding policies, the code window used to generate coded
packets must provide the potential for the client's knowledge space
to increase in the presence of packet losses. There are many ways
of accomplishing this goal ranging from schemes that code on a generation-by-generation
bases to schemes that code over the entire packet stream. While there
is no guarantee that the scheme proposed here is optimal, it does
lead to an admissible coding policy.

As a reminder, it is assumed that coded packets are used solely for
redundancy. If the path or network is error-free, coded packets will
not contribute to the knowledge space of the client. In addition,
it is assumed that coding occurs over a packet stream where the server
has limited to no knowledge of packets that will be sent in the future.
Therefore, any decisions regarding the code window management must
be made using information packets that have already been sent and
information from feedback that is at least $RTT$ seconds old. Before
an algorithm is proposed, the concept of a \textit{seen} packet from
\cite{sundararajan_network_2011} must be established.
\begin{defn}
The client is said to have \textit{seen} a packet $\boldsymbol{p}_{i}$
if it has enough information to compute a linear combination of the
form $\left(\boldsymbol{p}_{i}+\boldsymbol{q}\right)$ where $\boldsymbol{q}=\sum_{k>i}\alpha_{k}\boldsymbol{p}_{k}$
with $\alpha_{k}\in\mathbb{F}_{q}$ for all $k>i$. Therefore, $\boldsymbol{q}$
is a linear combination involving information packets with indices
larger than $i$.
\end{defn}
Define $\mathtt{seen}$ to be the index of the last \textit{seen}
information packet at the client that is composed of the set of all
consecutive \textit{seen} information packets. It is assumed that
the client informs the server of the value of $\mathtt{seen}$ through
feedback. Once the feedback has been received by the server, $\mathtt{seen}$
will be used to set the lower edge of the code window. The upper edge
of the code window will be managed based on the index of the last
transmitted uncoded information packet. This is summarized in Algorithm
\ref{alg:Coding-window-management}, which is executed by the server
and is agnostic to the path on which any one packet is transmitted.
\begin{algorithm}
\SetAlgoNoLine 
\SetAlgoNoEnd 
\DontPrintSemicolon 
\SetKwFor{ForEach}{for each}{do}{endfch} 
Initialize $\left(w_L,w_U\right)=\left( 0,0\right)$ and $j=0$\; 
\If{$\boldsymbol{p}_i$ transmitted uncoded \bf{and} $i>w_U$}{ $w_U\leftarrow i$ } 
\If{Feedback received \bf{and} $\mathtt{seen} > w_L$}{ $w_L \leftarrow \mathtt{seen}$ }

\caption{Code Window Management\label{alg:Coding-window-management}}
\end{algorithm}

Since $\mathtt{seen}$ is required to be the last \textit{seen} packet
out of the set of consecutive packets, the client will eventually
be able to decode given the transfer of enough degrees of freedom.
Furthermore, using \textit{seen} packets to manage the code window
helps to decrease the size of the coding/decoding buffers on the server/client
respectively.

\section{System Model\label{sec:System-Model}}

A time-slotted model is assumed where a single server-client pair
are communicating with each other over multiple parallel networks.
Similar to the last section, we denote this set of disjoint networks
as $\mathcal{P}$. Data is first placed into information packets $\boldsymbol{p}_{1},\boldsymbol{p}_{2}\ldots$
. These information packets are then used to generate coded packets
$\boldsymbol{c}_{1},\boldsymbol{c}_{2},\boldsymbol{c}_{3},\ldots$.
Depending on the coding policy, the server chooses to transmit either
an information packet or coded packet over one of the network paths.
The time it takes to transmit either type of packet is $t_{i}=\nicefrac{1}{r_{i}}$
seconds where $r_{i}$ is the transmission rate in packets/second
of network $i\in\mathcal{P}$. Furthermore, it takes each packet $d_{i}$
seconds to propagate through network $i$ (e.g., $RTT_{i}=t_{i}+2d_{i}$
on network $i$ assuming that the size of the feedback packet is sufficiently
small).

Delayed feedback is available to the server allowing it to estimate
each paths' independent and identically distributed (i.i.d.) packet
erasure rate $\epsilon_{i}$ and round-trip time $RTT_{i}$ (in seconds).
However, the server is unable to determine the cause of the packet
erasures (e.g., poor network conditions or congestion). Furthermore,
the server has knowledge of each network's transmission rate $r_{i}$,
which can either be determined from feedback obtained from the client
or from the size of the server's congestion window on any specific
path (e.g., $r_{i}=\nicefrac{cwnd_{i}}{RTT_{i}}$ ). This feedback
can also be used to communicate to the server the number of $dofs$
received by the client. While the following analysis assumes feedback
does not contain this information, numerical and simulated results
will use feedback to dynamically adjust the code rate depending on
the client's $dof$ deficit.

\section{Analysis of the In-Order Delivery Delay\label{sec:Analysis}}

Before proceeding, several assumptions are required to simplify the
analysis. First, it is assumed that coded packets are only sent over
a single network and the code rates conform to Corollary \ref{cor:For-the-caseAdmissibleCodingPolicyConstraints-SinglePath}.
The rate and packet erasure probability of the network used to send
coded packets will be referred to as $r_{c}$ and $\epsilon_{c}$
respectively. Second, packets transmitted over faster networks are
delayed so that they arrive in-order with packets transmitted over
slower networks. For example, assume that packets are transmitted
over two disjoint networks with propagation delays $d_{1}$ and $d_{2}$
where $d_{1}<d_{2}$. Packets transmitted over network $1$ will be
delayed an additional $d_{2}-d_{1}$ seconds. This assumption affects
the analysis by over-estimating the delay since there is a possibility
that packets transmitted over the faster networks can be delivered
in-order without waiting for a packet from the slower network. However,
the number of packets transmitted over the faster networks that can
be delivered without packets from the slower networks is relatively
small. Third, the coding window used for each coded packet contains
all transmitted information packets. This assumption is not necessary
if the code window management follows Algorithm \ref{alg:Coding-window-management}.
However, it does remove any ambiguity regarding the usefulness of
a received coded packet.

The in-order delivery delay for the code provided in Algorithm \ref{alg:Multi-path-coding-algorithm}
can be determined using a renewal-reward process based off of the
number of transmitted coded packets on path $p_{c}\in\mathcal{P}$.
More accurately, a renewal occurs whenever a received coded packet
results in a decoding event. This occurs whenever the number of received
coded packets is greater than or equal to the number of lost information
packets. Per Algorithm \ref{alg:Multi-path-coding-algorithm}, a coded
packet is transmitted every 
\begin{equation}
l_{c}=\frac{1}{1-c_{c}\left(\pi\right)}\label{eq:l_c}
\end{equation}
packets on path $p_{c}$. This results in the transmission of approximately
\begin{equation}
\alpha_{i}=\left(l_{c}-1\right)\frac{r_{i}}{r_{c}}\label{eq:alpha}
\end{equation}
information packets on each network $i\in\mathcal{P}$ for every transmitted
coded packet. Now consider a modified time-slotted model where each
time slot has duration $\nicefrac{l_{c}}{r_{c}}$. Define the sequence
$X_{1},X_{2},\ldots$, where $X_{n}=0,1,2,\ldots$ slots, to be the
i.i.d. inter-arrival times between decode events with first and second
moments $\mathbb{E}\left[X\right]$ and $\mathbb{E}\left[X^{2}\right]$
respectively. The arrival process is then a sequence of non-decreasing
random variables, or arrival epochs, $0\leq S_{1}\leq S_{2}\leq\cdots$
where the $n$th epoch $S_{n}=\sum_{i=1}^{n}X_{n}$.

In order to determine the distribution and moments of $X_{1},X_{2},\ldots$,
several additional random variables need to be defined. Let the random
variable $Y_{n,i}$, $i=1,2,\ldots$, be the number of lost packets
(both information and coded) between $S_{n-1}+\left(i-1\right)$ and
$S_{n-1}+i$ in the $n$th arrival epoch. The exact distribution is
the convolution of $\left|\mathcal{P}\right|$ binomial distributions
with parameters $\alpha_{i}$ and $\epsilon_{i}$ for each $i\in\mathcal{P}$.
In order to simplify the analysis, this distribution is approximated
by the following Poisson distribution: 
\begin{equation}
p_{Y_{n,i}}\left(y_{n,i}\right)=\frac{\lambda^{y_{n,i}}}{y_{n,i}!}e^{-\lambda},\quad y_{n,i}=0,1,\ldots,\label{eq:packet-loss-dist}
\end{equation}
 where 
\begin{equation}
\lambda=\epsilon_{c}+\sum_{i\in\mathcal{P}}\alpha_{i}\epsilon_{i}.\label{eq:lambda}
\end{equation}

If $Y_{n,1}=0$, the number of received packets between $S_{n-1}$
and $S_{n-1}+1$ is $1+\sum_{i\in\mathcal{P}}\alpha_{i}$ while only
$\sum_{i\in\mathcal{P}}\alpha_{i}$ packets were necessary to decode
(i.e., the coded packet is of no benefit and is dropped). Therefore,
$X_{n}=0$. However if $Y_{n,1}>0$, at least one packet was lost
which may prevent delivery. Therefore, the extra $dof$ obtained from
coded packets will help correct these erasures and eventually lead
to a decode event. As an example, consider the case when $Y_{n,1}=1$.
A single packet was lost, but enough $dofs$ were received to decode
all of the packets transmitted between $S_{n-1}$ and $S_{n-1}+1$.
Therefore, the inter-arrival time is $X_{n}=1$. Now consider the
case when $Y_{n,1}=2$ and $Y_{n,2}=0$. It is impossible for a renewal
to occur between at $S_{n-1}$ or $S_{n-1}+1$; however a renewal
does occur at $S_{n-1}+2$. This results in an inter-arrival time
$X_{n}=2$. Continuing on in this way, it becomes clear that a renewal
occurs the first time $Z$ that $\sum_{i=1}^{Z}Y_{n,i}\leq Z$.

In fact, $Z$ is a random variable and can be modeled as a M/D/1 queue
with a constant service time of $1$ packet per slot and an arrival
rate of $\lambda$ packets per slot. Define $Y=\sum_{i=1}^{Z}Y_{n,i}$,
then $Z$ conditioned on $Y$ has the following Borel-Tanner distribution
\cite{tanner61}: 
\begin{equation}
p_{Z|Y}\left(z|y\right)=\frac{yz^{z-y-1}\lambda^{z-y}e^{-z\lambda}}{\left(z-y\right)!},\label{eq:borel-tanner-dist}
\end{equation}
for $y=1,2,\ldots$ and $z=y,y+1,\ldots$. Equations (\ref{eq:packet-loss-dist})
and (\ref{eq:borel-tanner-dist}) can now be used to determine the
distribution of $X_{n}$ and its first two moments (proof is provided
in the appendix).
\begin{thm}
\label{thm:x_n}Let the number of packets lost in a single time-slot
be independent and identically distributed according to equation (\ref{eq:packet-loss-dist}).
The distribution of the time between decode events for all $\epsilon_{i}$
and $r_{i}$, $i\in\mathcal{P}$, that satisfy Corollary \ref{cor:For-the-caseAdmissibleCodingPolicyConstraints-SinglePath}
is 
\begin{equation}
p_{X_{n}}\left(x_{n}\right)=\begin{cases}
e^{-\lambda} & \text{for }x_{n}=0\\
\lambda e^{-\lambda} & \text{for }x_{n}=1\\
\frac{\left(x_{n}-1\right)^{x_{n}-2}}{x_{n}\left(x_{n}-2\right)!}\lambda^{x_{n}}e^{-x_{n}\lambda} & \text{for }x_{n}\geq2\\
0 & \text{otherwise},
\end{cases}
\end{equation}
with first and second moments 
\begin{equation}
\mathbb{E}\left[X\right]=\frac{\lambda}{1-\lambda}e^{-\lambda}
\end{equation}
 and
\begin{equation}
\mathbb{E}\left[X^{2}\right]=\left(\frac{1-\lambda+\lambda^{2}}{\left(1-\lambda\right)^{2}}\right)\mathbb{E}\left[X\right].
\end{equation}
\end{thm}
The first two moments of $X_{n}$ can now be used to determine the
the renewal-reward process that describes the in-order delivery delay.
Before this is done, the following lemma from \cite{GallagerStochasticProcesses}
is needed.
\begin{lem}
\label{lem:renewal-reward-process}Let $\left\{ R\left(t\right);t>0\right\} $
be a non-negative renewal-reward function for a renewal process with
expected inter-renewal time $\mathbb{E}\left[X\right]<\infty$. If
each $R_{n}$ is a random variable with $\mathbb{E}\left[R_{n}\right]<\infty$,
then with probability 1,
\begin{equation}
\lim_{t\rightarrow\infty}\frac{1}{t}\int_{t=0}^{t}R\left(\tau\right)d\tau=\frac{\mathbb{E}\left[R_{n}\right]}{\mathbb{E}\left[X\right]}.
\end{equation}
\end{lem}
Rather than defining the renewal reward function $R\left(t\right)$
and the renewal-reward process using the inter-arrival times $X_{n}$,
an estimate is considered where the inter-arrival times of this new
process are $W_{n}=\max\left(X_{n},1\right)$ (i.e., $W_{n}=1,2,\ldots$).
The distribution on $W_{n}$ and its first moment are defined in the
following.
\begin{figure}
\begin{centering}
\includegraphics[width=0.9\columnwidth]{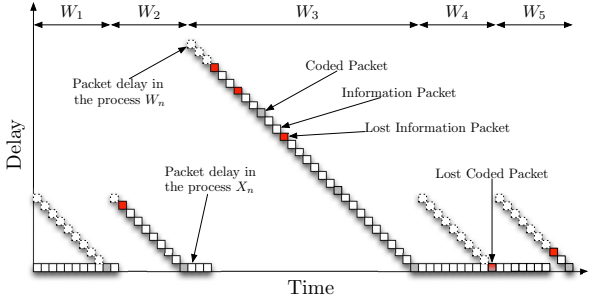}
\par\end{centering}
\caption{An example of the processes $X_{n}$ and $W_{n}$ and the reward function
$R\left(t\right)$ for a single path. Each box represents a transmitted
packet where the shaded boxes represent coded packets. The renewal
process $X_{n}$ is shown using only boxes with solid borders and
the process $W_{n}$ is shown using boxes with both solid and dotted
borders. Note that a renewal only occurs when a coded packet is received
by the decoder.\label{fig:reward_function}}
\end{figure}

\begin{cor}
\label{cor:w_n}Let the number of packets lost in a single time-slot
be independent and identically distributed according to equation (\ref{eq:packet-loss-dist})
and define $W_{n}=\max\left(X_{n},1\right)$. The the distribution
of inter-arrival times that satisfy Corollary \ref{cor:For-the-caseAdmissibleCodingPolicyConstraints-SinglePath}
is 
\begin{equation}
p_{W_{n}}\left(w_{n}\right)=\begin{cases}
\left(\lambda+1\right)e^{-\lambda} & \text{for }w_{n}=1\\
\frac{\left(w_{n}-1\right)^{w_{n}-2}}{w_{n}\left(w_{n}-2\right)!}\lambda^{w_{n}}e^{-w_{n}\lambda} & \text{for }w_{n}\geq2\\
0 & \text{otherwise},
\end{cases}
\end{equation}
where 
\begin{equation}
\mathbb{E}\left[W\right]=\frac{1}{\lambda}\mathbb{E}\left[X\right].
\end{equation}
\end{cor}
Figure \ref{fig:reward_function} provides a sample function of both
renewal processes $X_{n}$ and $W_{n}$. As a reminder, a renewal
is only possible when coded packets (shown using shaded boxes within
the figure) are received by the decoder. The reward function that
describes the in-order delay precisely is the area under the curve
shown for process $X_{n}$ (i.e., the blocks with solid borders).
However, the reward function $R\left(t\right)$ that describes the
delay for process $W_{n}$ (i.e., the union of boxes with dotted and
solid borders) is the one that is used. The in-order delivery delay
can now be determined by combining Lemma \ref{lem:renewal-reward-process},
Theorem \ref{thm:x_n}, and Corollary \ref{cor:w_n} (proof is provided
in the appendix).
\begin{thm}
\label{thm:sliding-window-delay}Consider the coding scheme described
by Algorithm \ref{alg:Multi-path-coding-algorithm} where redundant
packets are only transmitted on path $i_{c}\in\mathcal{P}$. With
probability one, the in-order delivery delay $\mathbb{E}\left[\Tau\right]$
is given by
\begin{align}
\mathbb{E}\left[\Tau\right]= & \frac{\lambda}{2r_{c}^{2}\left(1-\lambda\right)^{2}\sum_{j\in\mathcal{P}}r_{j}}\Biggl(r_{c}^{3}\left(l_{c}-1\right)\left(1-\lambda+\lambda^{2}\right)\nonumber \\
 & +\sum_{i\in\mathcal{P}\setminus i_{c}}\left(l_{c}r_{i}^{3}\left(1-\lambda+\lambda^{2}\right)-r_{i}^{2}r_{c}\left(1-\lambda\right)^{2}\right)\Biggr)\label{eq:sliding-window-delay}
\end{align}
\end{thm}

\section{Numerical and Simulation Results\label{sec:Numerical-and-Simulation}}

The expected in-order delivery delay $\mathbb{E}\left[\Tau\right]$
derived in the last section provides a useful approximation that can
be used to determine the performance of streaming codes operating
over multiple parallel network paths. Unlike the analysis in \cite{Karzand_2015},
the multi-path analysis is not technically an upper bound since approximations
were made with regards to $\alpha_{i}$'s and $\lambda$ (see (\ref{eq:alpha})
and (\ref{eq:lambda}) respectively). Regardless, the results presented
within this section show that the approximation is fairly good over
a range of network conditions. Furthermore, it should be noted that
the results presented within this section only compare the multi-path
delay with a single path. If a comparison with other coding approaches
such as generation/block codes is desired, the single path results
presented here along with the results presented in \cite{Karzand_2015}
can be used.
\begin{figure}
\begin{centering}
\includegraphics[width=1\columnwidth]{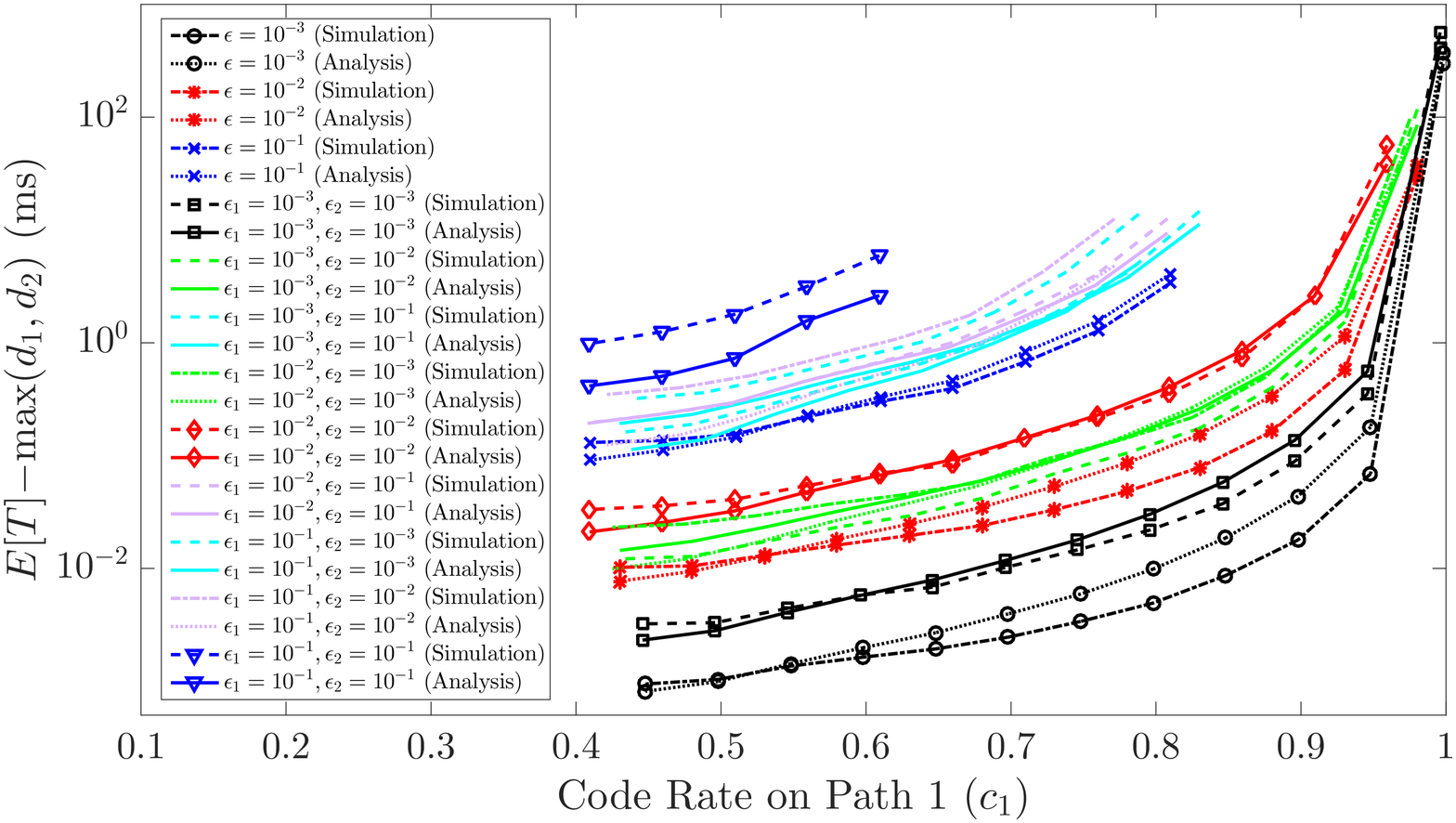}
\par\end{centering}
\caption{A comparison of the in-order delivery delay for a connection using
a single path (indicated using parameters without subscripts) and
one using two parallel paths (indicated using parameters with subscripts)
where $\nicefrac{r_{1}}{r_{2}}=\nicefrac{4}{3}$. These results only
show the case where coded packets are transmitted over a single path
$i_{c}=i_{1}$ with rate $r_{1}$.\label{fig:single-two-paths}}
\end{figure}

A comparison of the in-order delivery delay with simulated results
for the coding scheme presented earlier is provided in Figure \ref{fig:single-two-paths}
for both a session using a single path and a session using two paths.
In the case of the single path session, both coded and information
packets are transmitted on the path resulting in the following delay:
\begin{eqnarray}
\mathbb{E}\left[\Tau\right] & = & \frac{\lambda\left(l-1\right)\left(1-\lambda+\lambda^{2}\right)}{2\left(1-\lambda\right)^{2}}.\label{eq:sliding-window-delay-single-path}
\end{eqnarray}
In the case of the multi-path session, a single path with transmission
rate $r_{1}$ and packet erasure rate $\epsilon_{1}$ is used to transmit
all of the coded packets in addition to information packets. The second
path, which is only used to transmit only information packets, has
transmission rate $r_{2}$ and packet erasure rate $\epsilon_{2}$.
The figure demonstrates that the approximation developed earlier is
a fairly good measure of the true in-order delivery delay over a range
of code rates and packet erasure probabilities. Comparing the delay
for a session using a single path versus one using multiple parallel
paths, it is clear that the penalty for using multiple paths is not
that large. However, the potential benefits of diversifying the session
across paths can be significant. For example, the total throughput
of the session using two paths is almost twice that of the session
using only a single path. In addition, each additional path increases
the resiliency of the session. This can help ensure low delay when
network connections are transient or unreliable.

\begin{figure}
\begin{centering}
\includegraphics[width=1\columnwidth]{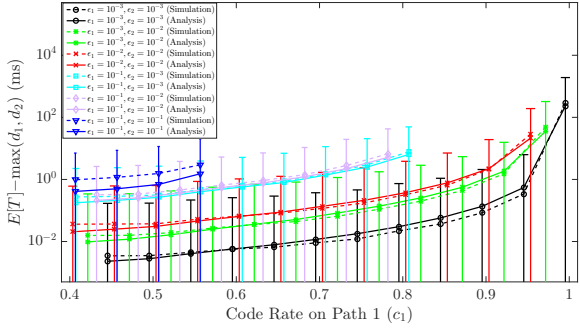}
\par\end{centering}
\caption{The in-order delivery delay for a connection using two paths where
$\nicefrac{r_{1}}{r_{2}}=\nicefrac{3}{4}$ where coded packets are
only transmitted over the path with packet erasure rate $\epsilon_{1}$.
The error bars show $\pm2\sigma_{\Tau}$.\label{fig:inorderdelivery_witherror} }
\end{figure}

This figure also shows that the delay is largely driven by the path
with the largest packet loss probability. For example, the delay for
paths with packet loss probabilities $\epsilon_{1}=10^{-2}$ and $\epsilon_{2}=10^{-3}$
is similar to the delay experienced by a single path with $\epsilon=10^{-2}$.
This is also the case when a path has packet loss probability $\epsilon_{i}=10^{-1}$.
Furthermore, the in-order delivery delay is not very sensitive to
changes in rate. This is illustrated by comparing Figure \ref{fig:single-two-paths}
with Figure \ref{fig:inorderdelivery_witherror}. The ratio of the
rates for each path in first figure is $\nicefrac{r_{1}}{r_{2}}=\nicefrac{4}{3}$,
while the ratio in the second figure is $\nicefrac{r_{1}}{r_{2}}=\nicefrac{3}{4}$.
Finally, Figure \ref{fig:inorderdelivery_witherror} also provides
information regarding the delay's variance. 

\section{Conclusions\label{sec:Conclusions}}

A streaming code that uses forward error correction to reduce in-order
delivery delay over multiple parallel networks was presented. This
included a discussion regarding the requirements for each paths' code
rate, in addition to methods to manage the generation of redundancy
through the use of a sliding code window. A simple analysis of the
code scheme was then developed that approximated the packet losses
on each of the paths using a Poisson distribution. Numerical results
were then presented showing that the analysis provides a good estimate
of the actual in-order delivery delay for a packet stream traversing
multiple parallels networks. These results illustrated that the path
with the largest packet erasure rate drives the overall in-order delivery
delay, and the delay is not very sensitive to changes in transmission
rates. Finally, the delay penalty for using multiple paths over a
single path was discussed. Numerical results helped show that this
delay penalty was insignificant with respect to the possible throughput
gains obtained by fusing two parallel networks together. 

\appendix{}
\begin{IEEEproof}
(Theorem \ref{thm:AdmissibleCodingPolicyConstraints}) A path with
code rate $c_{i}\left(\pi\right)$ and transmission rate $r_{i}$
packets/seconds results in a coded packet being generated every $\left(1-c_{i}\left(\pi\right)\right)r_{i}$
seconds. Therefore, the expected rate that coded packets arrive at
the client on path $i$ is $\left(1-\epsilon_{i}\right)\left(1-c_{i}\left(\pi\right)\right)r_{i}$
resulting in $\sum_{i\in\mathcal{P}}\left(1-\epsilon_{i}\right)\left(1-c_{i}\left(\pi\right)\right)r_{i}$
total coded packets/second. Now consider the case when each path is
treated as a separate session. The code rate on path $i$ must satisfy
$c_{i}<1-\epsilon_{i}$ in order to ensure the client's ability to
decode (i.e., the probability of a decoding error $Pr\{\mathcal{E}\}\rightarrow0$
as the file size increases for all $c_{i}<1-\epsilon_{i}$). Allowing
the code rate to be $c_{i}=1-\epsilon_{i}$, the expected rate at
which coded packets arrive at the client on path $i$ is then equal
to $\left(1-\epsilon_{i}\right)\epsilon_{i}r_{i}$ resulting in the
sum rate $\sum_{i\in\mathcal{P}}\left(1-\epsilon_{i}\right)\epsilon_{i}r_{i}$.
This produces the following bound: 
\begin{equation}
\sum_{i\in\mathcal{P}}\left(1-\epsilon_{i}\right)\left(1-c_{i}\left(\pi\right)\right)r_{i}>\sum_{i\in\mathcal{P}}\left(1-\epsilon_{i}\right)\epsilon_{i}r_{i}.\label{eq:admissable_policies-1-proof}
\end{equation}
Rearranging, we arrive at (\ref{eq:admissable_policies-1}). With
regard to (\ref{eq:admissable_policies-2}), it is obvious that $c_{i}\left(\pi\right)$
must satisfy $c_{i}\left(\pi\right)\in\left[0,1\right]$.
\end{IEEEproof}

\begin{IEEEproof}
(Theorem \ref{thm:x_n}) The inter-arrival time $X_{n}$ takes the
values of $x_{n}=0$ and $x_{n}=1$ if and only if $y_{n,1}=0$ with
probability $e^{-\lambda}$ and $y_{n,1}=1$ with probability $\lambda e^{-\lambda}$
respectively. For all $X_{n}\geq2$, we must have $y_{n,1}\geq2$.
This results in a decoding error in the first time-slot. Conditioning
on $Y_{n,1}$, we can use equation (\ref{eq:borel-tanner-dist}) to
find the probability for $X_{n}\geq2$ by setting $Z=X_{n}-1$ and
$Y=Y_{n,1}-1$:
\begin{align}
p_{X_{n}}\left(x_{n}\right) & =\sum_{y_{n,1}=2}^{x_{n}}p_{Y_{n,1}}\left(y_{n,1}\right)p_{Z|Y}\left(x_{n}-1|y_{n,1}-1\right)\\
 & =\sum_{y_{n,1}=2}^{x_{n}}\frac{\lambda^{y_{n,1}}e^{-\lambda}}{y_{n,1}!}\cdot\frac{\left(x_{n}-1\right)^{x_{n}-y_{n,1}-1}}{\left(x_{n}-y_{n,1}\right)!}\nonumber \\
 & \qquad\cdot\frac{\left(y_{n,1}-1\right)\lambda^{x_{n}-y_{n,1}}e^{-\left(x_{n}-1\right)\lambda}}{\left(x_{n}-y_{n,1}\right)!}\\
 & =\sum_{y_{n,1}=2}^{x_{n}}\frac{\left(y_{n,1}-1\right)\left(x_{n}-1\right)^{x_{n}-y_{n,1}-1}}{x_{1}!\left(z-x_{1}\right)!}\\
 & =\frac{\left(x_{n}-1\right)^{x_{n}-2}}{x_{n}\left(x_{n}-2\right)!}\lambda^{x_{n}}e^{-x_{n}\lambda}.
\end{align}
To determine the moments of $X_{n}$, first note that 
\begin{equation}
\sum_{x_{n}=2}^{\infty}\frac{\left(x_{n}-1\right)^{x_{n}-2}}{\left(x_{n}-2\right)!}\lambda^{x_{n}}e^{-x_{n}\lambda}=\mathbb{E}\left[X\right]-\lambda e^{-\lambda}
\end{equation}
and 
\begin{equation}
\sum_{x_{n}=2}^{\infty}\frac{x_{n}\left(x_{n}-1\right)^{x_{n}-2}}{\left(x_{n}-2\right)!}\lambda^{x_{n}}e^{-x_{n}\lambda}=\mathbb{E}\left[X^{2}\right]-\lambda e^{-\lambda}.
\end{equation}
We can then take the first and second derivatives of 
\begin{equation}
\sum_{x_{n}=0}^{\infty}p_{X_{n}}\left(x_{n}\right)=1,
\end{equation}
i.e., 
\begin{equation}
\frac{\partial^{i}}{\partial\lambda^{i}}\left(e^{-\lambda}+\lambda e^{-\lambda}+\sum_{x_{n}=2}^{\infty}\frac{\left(x_{n}-1\right)^{x_{n}-2}}{x_{n}\left(x_{n}-2\right)!}\lambda^{x_{n}}e^{-x_{n}\lambda}\right)=0
\end{equation}
for $i=\left\{ 1,2\right\} $, to find $\mathbb{E}\left[X\right]$
and $\mathbb{E}\left[X^{2}\right]$ respectively. For both $\mathbb{E}\left[X\right]<\infty$
and $\mathbb{E}\left[X^{2}\right]<\infty$, the rate of packet loss
across all paths must be $\lambda<1$. This corresponds with Corollary
\ref{cor:For-the-caseAdmissibleCodingPolicyConstraints-SinglePath}
after substituting in equations (\ref{eq:l_c}) and (\ref{eq:alpha})
when $l_{c}\epsilon_{i}\approx1$ $\forall i\in\mathcal{P}$.
\end{IEEEproof}

\begin{IEEEproof}
(Theorem \ref{thm:sliding-window-delay}) The renewal-reward function
to determine the delay experienced by an information packet transmitted
on path $P\in\mathcal{P}$ is similar to the residual life of the
process with some modifications. Define $R_{n}$ given $W_{n}$ to
be the sum delay of all information packets on path $P$: 
\begin{eqnarray}
R_{n} & = & \begin{cases}
\sum_{k=1}^{W_{n}l_{c}}k-l_{c}\sum_{k=1}^{W_{n}}k & \mbox{for }P=i_{c}\\
\frac{r_{c}}{r_{i}}\sum_{k=0}^{\frac{W_{n}l_{c}r_{i}}{r_{c}}-1}k & \mbox{for }P\neq i_{c}
\end{cases}\label{eq:R_n_W}
\end{eqnarray}
Taking the expectation of $R_{n}$, we obtain the following
\begin{align}
\mathbb{E}\left[R_{n}\right] & =\sum_{i\in\mathcal{P}}\sum_{w_{n}=1}^{\infty}\mathbb{E}\left[R_{n}|P=i,W_{n}=w_{n}\right]p_{P}\left(i\right)p_{W_{n}}\left(w_{n}\right)\\
 & =\frac{1}{\sum_{j\in\mathcal{P}}r_{j}}\sum_{w_{n}=1}^{\infty}\Biggl(r_{c}\mathbb{E}\left[\sum_{k=1}^{w_{n}l_{c}}k-l_{c}\sum_{k=1}^{w_{n}}k\right]\nonumber \\
 & \qquad\qquad+\sum_{i\in\mathcal{P}\setminus i_{c}}r_{i}\mathbb{E}\left[\frac{r_{c}}{r_{i}}\sum_{k=0}^{\frac{w_{n}l_{c}r_{i}}{r_{c}}-1}k\right]\Biggr)p_{W_{n}}\left(w_{n}\right).\label{eq:R_n_w_theorem}
\end{align}
Substituting $x_{n}$ for $w_{n}$ from Corollary \ref{cor:w_n},
\begin{align}
\mathbb{E}\left[R_{n}\right] & =\frac{1}{\sum_{j\in\mathcal{P}}r_{j}}\sum_{x_{n}=0}^{\infty}\Biggl(r_{c}\mathbb{E}\left[\sum_{k=1}^{\max\left(x_{n},1\right)l_{c}}k-l_{c}\sum_{k=1}^{\max\left(x_{n},1\right)}k\right]\nonumber \\
 & \qquad+\sum_{i\in\mathcal{P}\setminus i_{c}}r_{i}\mathbb{E}\left[\frac{r_{c}}{r_{i}}\sum_{k=0}^{\frac{\max\left(x_{n},1\right)l_{c}r_{i}}{r_{c}}-1}k\right]\Biggr)p_{X_{n}}\left(x_{n}\right)\\
 & =\frac{1}{\sum_{j\in\mathcal{P}}r_{j}}\Biggl(\frac{l_{c}r_{c}}{2}\left(l_{c}-1\right)\mathbb{E}\left[X^{2}\right]\nonumber \\
 & \qquad+\sum_{i\in\mathcal{P}\setminus i_{c}}\frac{l_{c}r_{i}^{2}}{2r_{c}^{2}}\left(l_{c}r_{i}\mathbb{E}\left[X^{2}\right]-r_{c}\mathbb{E}\left[X\right]\right)\Biggr).
\end{align}
Since both $\mathbb{E}\left[X\right]<\infty$ and $\mathbb{E}\left[X^{2}\right]<\infty$,
the expectation $\mathbb{E}\left[R_{n}\right]<\infty$ and Lemma \ref{lem:renewal-reward-process}
can be applied. Keeping in mind that every time-slot in the process
defined by $W_{n}$ is divided into $l_{c}$ smaller time-slots, the
expected in-order delivery delay is 
\begin{align}
\mathbb{E}\left[\Tau\right] & =\lim_{t\rightarrow\infty}\frac{1}{l_{c}t}\int_{0}^{t}R\left(\tau\right)d\tau\\
 & =\frac{\mathbb{E}\left[R_{n}\right]}{l_{c}\mathbb{E}\left[W\right]}\\
 & =\frac{\lambda}{\mathbb{E}\left[X\right]\sum_{j\in\mathcal{P}}r_{j}}\Biggl(\frac{r_{c}}{2}\left(l_{c}-1\right)\mathbb{E}\left[X^{2}\right]\nonumber \\
 & \qquad+\sum_{i\in\mathcal{P}\setminus i_{c}}\frac{r_{i}^{2}}{2r_{c}^{2}}\left(l_{c}r_{i}\mathbb{E}\left[X^{2}\right]-r_{c}\mathbb{E}\left[X\right]\right)\Biggr)\\
 & =\frac{\lambda}{2r_{c}^{2}\left(1-\lambda\right)^{2}\sum_{j\in\mathcal{P}}r_{j}}\Biggl(r_{c}^{3}\left(l_{c}-1\right)\left(1-\lambda+\lambda^{2}\right)\nonumber \\
 & +\sum_{i\in\mathcal{P}\setminus i_{c}}\left(l_{c}r_{i}^{3}\left(1-\lambda+\lambda^{2}\right)-r_{i}^{2}r_{c}\left(1-\lambda\right)^{2}\right)\Biggr)
\end{align}
\end{IEEEproof}
\bibliographystyle{ieeetr}
\bibliography{IEEEfull,MultipathStreamingCode}

\begin{thebibliography}{10}

\bibitem{Karzand_2015}
M.~Karzand, D.~Leith, J.~Cloud, and M.~M\'{e}dard, ``{Low Delay Random Linear
  Coding Over a Stream},'' {\em CoRR}, vol.~abs/1509.00167, 2015.

\bibitem{Cloud_15}
J.~Cloud, D.~Leith, and M.~M{\'e}dard, ``{A Coded Generalization of Selective
  Repeat ARQ},'' in {\em IEEE Conference on Computer Communications (INFOCOM)},
  pp.~2155--2163, April 2015.

\bibitem{joshi_playback_2012}
G.~Joshi, Y.~Kochman, and G.~W. Wornell, ``{On Playback Delay in Streaming
  Communication},'' in {\em IEEE International Symposium on Information Theory
  Proceedings (ISIT)}, pp.~2856--2860, July 2012.

\bibitem{toemoeskoezi_delay_2014}
M.~T\"{o}m\"{o}sk\"{o}zi, F.~H. Fitzek, F.~H. Fitzek, D.~E. Lucani, M.~V.
  Pedersen, and P.~Seeling, ``{On the Delay Characteristics for Point-to-Point
  Links using Random Linear Network Coding with On-the-Fly Coding
  Capabilities},'' in {\em 20th European Wireless Conference; Proceedings of
  European Wireless}, pp.~1--6, May 2014.

\bibitem{Han_2008}
S.~Han, Z.~Zhong, H.~Li, G.~Chen, E.~Chan, and A.~K. Mok, ``{Coding-Aware
  Multi-path Routing in Multi-Hop Wireless Networks},'' in {\em IPCCC},
  pp.~93--100, Dec 2008.

\bibitem{Ronasi_2011}
K.~Ronasi, A.~H. Mohsenian-Rad, V.~W.~S. Wong, S.~Gopalakrishnan, and
  R.~Schober, ``{Delay-Throughput Enhancement in Wireless Networks With
  Multipath Routing and Channel Coding},'' {\em IEEE Transactions on Vehicular
  Technology}, vol.~60, pp.~1116--1123, March 2011.

\bibitem{Nguyen_2004}
V.~T. Nguyen, E.~C. Chang, and W.~T. Ooi, ``{Layered coding with good
  allocation outperforms multiple description coding over multiple paths},'' in
  {\em ICME}, vol.~2, pp.~1067--1070 Vol.2, June 2004.

\bibitem{GarciaSaavedra_2015}
A.~Garcia{-}Saavedra, M.~Karzand, and D.~J. Leith, ``{Low Delay Random Linear
  Coding and Scheduling Over Multiple Interfaces},'' {\em CoRR},
  vol.~abs/1507.08499, 2015.

\bibitem{ho_random_2006}
T.~Ho, M.~M\'{e}dard, R.~Koetter, D.~Karger, M.~Effros, J.~Shi, and B.~Leong,
  ``{A Random Linear Network Coding Approach to Multicast},'' {\em {IEEE}
  Transactions on Information Theory}, vol.~52, pp.~4413--4430, October 2006.

\bibitem{sundararajan_network_2011}
J.~K. Sundararajan, D.~Shah, M.~M\'{e}dard, S.~Jakubczak, M.~Mitzenmacher, and
  J.~Barros, ``{Network Coding Meets {TCP}: Theory and Implementation},'' {\em
  Proceedings of the IEEE}, vol.~99, pp.~490--512, March 2011.

\bibitem{tanner61}
J.~C. Tanner, ``A derivation of the borel distribution,'' {\em Biometrika},
  vol.~48, no.~1/2, pp.~pp. 222--224, 1961.

\bibitem{GallagerStochasticProcesses}
R.~G. Gallager, {\em Stochastic Processes: Theory for Applications}.
\newblock New York, NY: Cambridge University Press, 2013.

\end{thebibliography}

\end{document}